\documentclass[conference]{IEEEtran}
\IEEEoverridecommandlockouts
\usepackage{cite}
\usepackage{amsmath,amssymb,amsfonts}
\usepackage{algorithmic}
\usepackage{graphicx}
\usepackage{textcomp}
\usepackage{xcolor}
\usepackage{float}
\def\BibTeX{{\rm B\kern-.05em{\sc i\kern-.025em b}\kern-.08em
    T\kern-.1667em\lower.7ex\hbox{E}\kern-.125emX}}
\begin{document}

\title{Can Social Robots Effectively Elicit Curiosity in STEM Topics from K-1 Students During Oral Assessments?}

\makeatletter
\newcommand{\linebreakand}{%
  \end{@IEEEauthorhalign}
  \hfill\mbox{}\par
  \mbox{}\hfill\begin{@IEEEauthorhalign}
}
\makeatother

\author{\IEEEauthorblockN{Alexander Johnson}
\IEEEauthorblockA{\textit{Electrical and Computer Engineering} \\
\textit{University of California, Los Angeles}\\
Los Angeles, USA \\
ajohnson49@ucla.edu}
\and
\IEEEauthorblockN{Alejandra Martin}
\IEEEauthorblockA{\textit{Department of Education} \\
\textit{University of California, Los Angeles}\\
Los Angeles, USA \\
alemartin@ucla.edu }
\and
\IEEEauthorblockN{Marlen Quintero}
\IEEEauthorblockA{\textit{Department of Education} \\
\textit{University of California, Los Angeles}\\
Los Angeles, USA \\
mquint30@g.ucla.edu}
\linebreakand

\IEEEauthorblockN{Alison Bailey}
\IEEEauthorblockA{\textit{Department of Education} \\
\textit{University of California, Los Angeles}\\
Los Angeles, USA \\
abailey@gseis.ucla.edu }

\and

\IEEEauthorblockN{Abeer Alwan}
\IEEEauthorblockA{\textit{Electrical and Computer Engineering} \\
\textit{University of California, Los Angeles}\\
Los Angeles, USA \\
alwan@ee.ucla.edu }

}

\maketitle

\begin{abstract}

This paper presents the results of a pilot study that introduces social robots into kindergarten and first-grade classroom tasks.  This study aims to understand 1) how effective social robots are in administering educational activities and assessments, and 2) if these interactions with social robots can serve as a gateway into learning about robotics and STEM for young children.  We administered a commonly-used assessment (GFTA3) of speech production using a social robot and compared the quality of recorded responses to those obtained with a human assessor.  In a comparison done between 40 children, we found no significant differences in the student responses between the two conditions over the three metrics used: word repetition accuracy, number of times additional help was needed, and similarity of prosody to the assessor.  We also found that interactions with the robot were successfully able to stimulate curiosity in robotics, and therefore STEM, from a large number of the 164 student participants.

\end{abstract}


\section{Introduction}

Social robots have proven to be an effective aid to early childhood language acquisition \cite{b1}. Their welcoming designs and expressive movements make them engaging for children to speak and play with \cite{b2}. Several studies have shown that children experience psychological and educational benefits by spending time with social robots (\cite{b3,b4}) including students' improved academic and social outcomes and student engagement.  The recent success of these devices with children may also imply that their introduction to educational settings will make children more curious about them and, more broadly, robotics and STEM.  Previous studies such as \cite{b5} and \cite{b6} show the effectiveness of robots as teaching tools in STEM lessons through collected user surveys, and \cite{b7} introduces the design for a conversational robotic interface that engages primary school children in STEM discussions.  However, further work is needed to explore how robots can be effectively integrated into existing lessons and tasks for young children rather than creating new robot-centered lessons.  In addition, more free-response feedback is needed in order to determine what aspects of human-robot interaction pique young children’s curiosity in robotics and could be used to motivate them to later pursue STEM education.
Inspired by previous work using voice-enabled social robots to encourage children’s oral language development through storytelling (\cite{b8,b9}), we apply social robots to the critical task of conducting assessments of language and early literacy-related skills in children.  Such assessments are particularly important for kindergarten and first-grade children, as they are beginning to learn to read, and any oral language delays should be addressed before they negatively impact early literacy development.  Children’s early language abilities are also indicators of their later aptitude in other subjects including writing, math, and science.  Public schools typically do not have sufficient support from speech-language pathologists, reading specialists, or literacy coaches to conduct detailed, dialect-appropriate assessments of every child’s oral language abilities on a routine basis, depriving some children in need of extra language support or intervention from such resources.  A social robot enabled with automatic speech recognition (ASR) technology may be used to conduct some of these intensive assessments with children without the need for a trained language specialist.  The presence of the social robot during these assessments can also become a seamless way to introduce children to robotics and STEM.  However, before these systems are implemented, it must be determined whether or not the presence of a robot has a profound effect on the children’s speech.  A person’s tone, prosody, and word choice can vary with whomever they are speaking.  Therefore, further investigation is needed to verify whether or not children’s speech productions differ significantly in the presence of a voice-enabled social robot, compared to a human oral assessment administrator, in a way that might hinder obtaining accurate assessments of speech production.  In this work, we investigate 1) how effective social robots are in the direct assessment of children's speech production (phonological processing skills), and 2) if social robots can be used to promote young children’s curiosity in robotics and STEM during necessary oral language assessments.


In order to investigate these questions, we conduct common oral language assessments with kindergarten and first grade children in two scenarios:  In one case, the assessments are administered by a human assessor as typically conducted.  In the other case, assessments are administered by, JIBO \footnote{`Jibo Robot - He can't wait to meet you," Boston, MA, 2017. [Online]. Available: https://www.jibo.com}, a social robot designed for interacting with children (pictured in Figure 1), while the human assessor watches and intervenes only as necessary.  We selected children of these grade levels because they are at the critical age range for language development and literacy acquisition (\cite{b10, b11}).  To identify any significant differences in speech production between children in the two cases that may affect assessment scores, we investigate the students' changes in behavior, prosody, and accuracy of word repetition during the assessments across the two scenarios for both grade levels.  We additionally documented the children's responses to interacting with JIBO that indicate curiosity in robotics or may be used to further interest in STEM-related topics.

\section{Methods}

\subsection{Participants}

A sample of 164 kindergarten and first-grade students, consisting of 53 kindergarteners and 111 first graders from a Southern California elementary school, were recruited for the study.  Selection criteria included parental consent and completion of all tasks in the study.  All of the students were English speaking with some reporting additional language exposure (most often Spanish) at home.

\subsection{Recording}

Interactions with each student were recorded with a Logitech C920 Webcam microphone with a sampling rate of 48kHz.  Each student was approximately two feet from the test administrator (either JIBO or the human assessor), and the microphone was placed equidistantly between them.  Recordings took place in an empty office at the school site during school hours to simulate a realistic environment in which JIBO could be used. 

\subsection{Experiment}

Data collection followed the protocol described in  (\cite{b12, b13, b14}). The children\textquotesingle s speech samples were collected using Goldman-Fristoe Test of Articulation-3 (GFTA-3) \cite{b15} sounds in sentences protocol.  Each student was first read a story appropriate for their grade level.  The kindergarten students were read a story containing 20 simple sentences and the first-grade students were read a story containing 15 sentences mixing both simple and compound structures.  As a fictitious example (the actual assessment material is under copyright restrictions), a sentence such as, ``The cat was fat, fuzzy, and orange," may have appeared in the story for first-graders.  After the story was read once in its entirety, the story was repeated sentence by sentence, and the student was asked  to repeat each sentence back to the test administerer.  Ten kindergarten students and ten first-grade students were administered the test by a human assessor, and the others were administered the test by JIBO.  Both gave the same prompts to the children.  In giving each prompt, JIBO played a corresponding audio file containing a recording of an adult woman whose voice was pitch shifted up to match the pitch of a child and slowed down.  For sessions in which the child was given the prompts by JIBO, a human assessor was present and intervened if the child was unable to complete the task with only the social robot\textquotesingle s instructions.  In sessions in which the child was given the prompt by a human assessor, they gave the prompt to the child a second time if they were unable to repeat the sentence after hearing it once.

Any questions or comments that the child made referring to JIBO before, during, or after the assessment were documented and categorized by their different characteristics.  

\begin{figure}[h]
\centering
\includegraphics[width=0.5\textwidth]{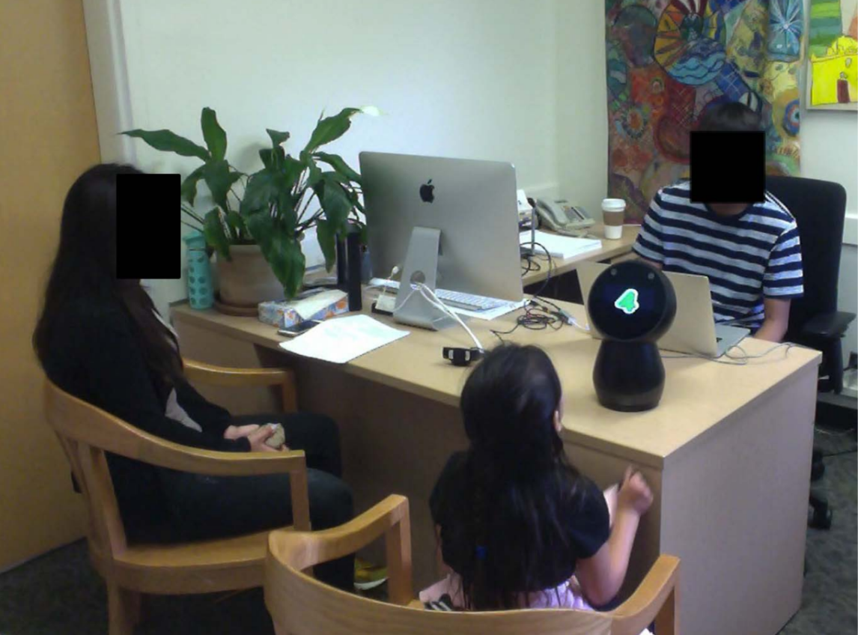}
\caption{A session in which JIBO administers the GFTA-3 (sounds in sentences assessment) to a student.}
\end{figure}

\subsection{Assessment Quality Evaluation}

The 10 kindergarteners and 10 first-graders who completed the assessment with only the human assessor were compared to a group of 10 kindergarteners and 10 first-graders who completed the same assessment administered by the social robot.  The robot-assessed children in the comparison group were selected from the total participant pool on the basis that they were assessed in the same time period and overseen by same human assessor as the students in the human-assessed children.  The following metrics were calculated for comparison between the two conditions:

\begin{enumerate}
  \item Accuracy: each sentence that the students were asked to repeat was scored out of the total number of words in the sentence.  One point was given for each word that the child repeated correctly in the correct order.
  \item Need for additional prompting: we noted each time that a child required additional prompting to repeat a sentence and noted the reason for the additional prompting.  If the child was silent for at least 3 seconds after given the prompt, the reason was noted as ``reticence."  If the child forgot the prompt or did not understand what to say, the reason is given as ``needs repetition."  If the child said something unrelated to the prompt then the reason was listed as ``distracted."
\item Pitch correlation: An additional question is whether students attempt to follow the administrator's pitch and rhythm during the assessment.  While not directly related to speech production assessment quality, differences in how the children incorporate the human and the robot's prosody into their repetition of the prompt may indicate how engaged the children were with the assessor.  For each prompt, the pitch contour of both the administerer (JIBO or the human assessor) uttering the prompt and the child\textquotesingle s voice while repeating it were extracted using Praat\textquotesingle s \cite{b16} pitch tracking algorithm with parameters set to best measure a child\textquotesingle s pitch.  The Praat pitch contours were manually corrected if any errors occurred in the pitch tracking algorithm.  Then the Pearson correlation coefficient between the child\textquotesingle s pitch contour and administrator\textquotesingle s (either the robot or human assessor) pitch contour was calculated and taken as a measure of the perceptual similarity of the two as in \cite{b17}.
\end{enumerate}

Sentence word accuracy scores were scored manually by the researchers who are experienced in evaluating children\textquotesingle s speech.  Reasons for the child needing additional prompting were also noted manually following the above criteria.

\subsection{Student Responses}

The students were allowed to interact with the robot and speak freely before, during, and after the assessment.  Any questions or comments made by students about JIBO that may be used to grow interest in robotics or STEM were characterized by the following five categories:

\begin{enumerate}
\item Mechanical: Referring to how robots work or asking about fundamentals of robotics or other STEM concepts
\item Functional: Referring to why JIBO looks and acts as it does or asking about engineering design choices
\item Relational: Relating JIBO to something that the student has seen previously
\item Personifying: Assigning human characteristics to JIBO
\item Hypothetical: Exploratory questions about JIBO or related STEM concepts 
\end{enumerate}

Questions/comments could by characterized by more than one category.  Some examples of children's comments are given in Table I.

\begin{table}[H]
\begin{tabular}{|| c | p{60mm} ||} \hline
Category  & Examples \\ \hline
Mechanical & ``You program him to do that?", \newline ``He didn't get rusty" \\ \hline
Functional & ``Why is JIBO have one eye?", \newline ``Why he dance like this?"  \\ \hline
Relational & ``I have a robot at my home", \newline ``I know robot [that] is more strong"  \\ \hline
Personifying & ``Now, JIBO is looking at me", \newline ``You miss your mommy, JIBO or not?" \\ \hline
Hypothetical & ``But what if you flip JIBO backwards?", \newline ``Can JIBO do a handstand?" \\ \hline
\end{tabular}
\\
\caption{\label{tab:table-name} Some examples of comments or questions from children when interacting with the social robot.}
\end{table}

\section{Results}

In this section, we present the results of the experiment.  Subsections A-C show the results comparing 40 children (10 from each category: kindergarten or first-grade, assessed by the human or JIBO). Section D summarizes our observations of the 144 children assessed by JIBO (164 less the 20 assessed by only a human).

\subsection{Accuracy of Repetition}

Table II shows the percent word accuracy of the repetition task for children in each group.  The maximum score, minimum score, and standard deviation for the group are also given.

\begin{table}[htb]
\begin{tabular}{|| c | c | c | c | c ||} \hline
 Administration & Mean & STDV & Min & Max \\ \hline
Kindergarteners with JIBO & 95.08 & 12.62 & 86.15 & 100 \\ \hline
Kindergarteners w/o JIBO & 97.85 & 7.0 & 93.7 & 100 \\ \hline
First Graders with JIBO &91.91  & 17.0 & 63.67 & 100 \\ \hline
First Graders w/o JIBO & 93.13 & 13.02 & 81.65 & 100 \\ \hline
\end{tabular}
\\
\caption{\label{tab:table-name}Average percent of words repeated correctly (accuracy) for a group with percent standard deviation (STDV) and minimum and maximum percent scores of every child in the category}
\end{table}

\begin{table*}[t]
\begin{tabular}{|| c | c | c | c | c | c | c ||} \hline
Administration  & Reticent & Reticent at beginning & Max Reticent & Needs Repetition & Max Needs Rep. & Distracted  \\ \hline
Kindergarteners with JIBO & 8 & 5 & 3 & 4 & 2 & 1 \\ \hline
Kindergarteners w/o JIBO & 1 & 0 & 1 & 0 & 0 & 2 \\ \hline
First Graders with JIBO & 3 & 2 & 2  & 13 & 7 & 0 \\ \hline
First Graders w/o JIBO & 2 & 0 & 1 & 6 & 2 & 1 \\ \hline
\end{tabular}
\\
\caption{\label{tab:table-name}Number of times a child (10 children in each category, resulting in a total of 40 students) needed additional prompting from the examiner in order to repeat the prompt because of: A)  Reticence the child taking 3 or more seconds to respond after being given a prompt, B)  Reticent at beginning notes the number of reticent instances that occurred at the beginning of the session, C) Max Reticent is the largest number of times a single child showed signs of being reticent in a session.  Needs repetition is defined as the number of instances in which a child was unable to completely say the prompt without being reminded of some or all words.  Max Needs Rep. is the largest number of times that a single child needed some or all of a prompt to be repeated.  Distracted is defined as the number of times a child gave an unrelated answer.}
\end{table*}

\subsection{Additional Prompting}
Table III gives the number of times students in the two groups required additional prompting to be able to complete the exercise and the reason for the need for additional prompting.

\subsection{Prosody}

Average correlation between the assessor and students' pitch contours when uttering the same prompt are shown in Table IV for each group.  The percentage of repetitions spoken by students that are significantly correlated in pitch contour to the prompt given by the assessor are also given in this Table.

\begin{table}[H]
\begin{tabular}{|| c | c | c | c||} \hline
Administration & Mean Corr Coef & STDV & \% Sig  \\ \hline
Kindergarteners with JIBO & 0.191  & 0.131 & 67.5  \\ \hline
Kindergarteners w/o JIBO  & 0.258 & 0.151 & 72.5  \\ \hline
First Graders with JIBO & 0.145 & 0.108 & 54.0  \\ \hline
First Graders w/o JIBO & 0.179 & 0.143 & 56.7 \\ \hline
\end{tabular}
\\
\caption{\label{tab:table-name} Pearson\textquotesingle s correlation coefficient between the examiner\textquotesingle s pitch contour and the child\textquotesingle s pitch contour upon repeating the examiner\textquotesingle s utterance.  The percentage of child utterance whose pitch contour correlated significantly ($p<0.05$) to that of the original prompt is also given.}
\end{table}


\subsection{Scientific Curiosity}

In total, 43 kindergarteners and 101 first-graders were assessed using JIBO. Table V shows the percentage of these children in each grade level who made a comment or question in the given category.  In total, approximately 45\% of the students made comments or questions about JIBO pertaining to at least one of the categories.

\begin{table}[H]
\begin{tabular}{|| c | c | c | c | c | c | c||} \hline
\% Students & Func. & Mech. & Rel. & Person. & Hypoth. & Total \\ \hline
Kinder. & 31.81 & 20.45 & 0.00 & 29.54 & 4.54 & 53.2 \\ \hline
First. & 10.89 & 17.82 & 3.96 & 19.80 & 3.96 & 39.1\\ \hline
\end{tabular}
\\
\caption{\label{tab:table-name} Percent of students who made a question or comment by given type (Functional, Mechanical, Relational, Personifying, or Hypothetical) while interacting with JIBO.  The same child could be counted in multiple categories if their questions/comments fell into more than one.  The total percentage of students who made a question or comment is also given.}
\end{table}

\section{Discussion}

\subsection{Evaluation of Speech Assessment Quality}

As shown in Table II, in the repetition task, the students\textquotesingle \ performance was not significantly affected by the presence of the social robot, JIBO.  The table shows that when repeating words to the examiner, neither the kindergarten students ($p>0.1$) nor the first-grade students ($p>0.05$) are significantly worse at repeating words with JIBO than they are with a human.  Although the mean sentence score was about 2\% higher when  the test was administered by a human assessor for both the kindergarten and first-grade students, this number is not statistically significant ($p>0.05$).  There is however a significantly larger standard deviation in the sessions administered by JIBO (note the differences between the minimum and maximum sentence scores).  

Table III shows that kindergarteners are more reticent when the test is administered by JIBO than by a human assessor.  However, this reticence typically only occurs at the beginning of the session and disappears as the students are exposed more to the social robot.  A possible explanation for this finding is that at this early stage of the activity, students may have been adapting to the activity.  This trend in reticence does not extend to the first-grade students.  Most of the first-grade students did need sentences to be repeated more often by the human facilitator present when the test was administered by JIBO.  It is worth noting that the first-grade student who needed intervention most often when using JIBO predominately spoke a language other than English at home.  The prompts that needed repetition were most often those containing longer sentences with compound structures. 

Table IV implies that the pitch contours of the students\textquotesingle \ repetition of the prompt more closely match those of the prompt given by the human assessor. This means that while repeating sentences said by the assessor, the students' changes in pitch and rhythm matched the human assessor more closely than the robot.  This may be due to the robot having less natural sounding prosody, making it more difficult to match. However, these correlations are not significant, and we would need a larger sample size to definitively state that the use of a robot affects whether or not the students follow the speech prosody of the human assessor more.  It is important to note that the correlations for first graders are lower than those of the kindergarteners.  This is likely due to the fact that the prompts read to the first graders were longer and contained more complex sentence structures, requiring the students to turn their attention more towards remembering the words than matching the prosody given.


\subsection{Motivating Scientific Curiosity}
Almost half of the students made inquiries as to how robots work, why they are designed in the way that they are, and other ideas that are conducive to furthering children's interest in STEM.  The kindergarteners most commonly asked questions related to how JIBO is able to move, speak, and show images.  That is, their questions typically asked for explanations to observed phenomenon.  This implies that these interactions can be used to motivate kindergarten-level lessons on more visual concepts in robotics and STEM like motors, cameras, and programming fundamentals.  The first-graders also asked a large number of questions of this nature, albeit fewer on average than the kindergarteners, while also asking personifying and hypothetical questions that delve past just what they can directly observe in the robot, such as, ``Where does JIBO sleep?" and ``Can you make him do [a different type of movement]?" These questions may be used to motivate further interest into engineering concepts such as biological inspiration for design and user-centered design.  We also note that only the first graders related JIBO to other robots that they had seen previously.  This may be because the younger students have had less exposure to other instances of robots.  

We observed that students became excited to participate in the assessment when they learned that they would be working with a robot.  This experience was, for most, if not all, of the students, their first interaction with a social robot.  From the students' excitement and frequency of questions asked, we believe that JIBO can be used to prime young students for robotic and STEM-centered lessons, possibly motivating them to further pursue them in secondary and post-secondary education.  We believe that many of the students who did not make comments or questions about JIBO were simply shy, as our work shows that even the students working with only the human assessor still displayed a large amount of reticence throughout the assessment.  Further interactions with the robot to build familiarity may reduce that reticence over time and lead to additional questions from students in the future.

\section{Limitations}

Comparing our study with others on robot-directed speech (\cite{b18,b19}) using a power analysis, we noted that that our results on children's human-directed and robot-directed speech are underpowered.  However, our findings demonstrate strong tendencies and suggest directions for future research.
In addition, variability might have been reduced if the same children had completed the assessment under both assessment conditions (with or without JIBO), giving different, equally difficult prompts in each condition.  However, in the current study, we had only a  short duration in each session to work with the children, making it infeasible to give one child both assessment conditions within a session. We instead balanced the pools for each condition as effectively as possible within the constraints.

\section{Conclusions}

Our pilot results show no detriment in the students' speech production performances as a result of working
with the social robot. Students were sometimes more hesitant to talk to the robot at first but became
more willing over the course of the session. The time spent with the social robot also appears to have
elicited questions from the students that can be used to grow their interest in this STEM domain. Future work includes measuring students' curiosity and speech differences in a longitudinal study to investigate how these factors change with time.

\section{Acknowledgements}
This work is supported in part by the NSF.

\vspace{12pt}

\end{document}